# AN OPPORTUNISTIC AODV ROUTING SCHEME: A COGNITIVE MOBILE AGENTS APPROACH

Seema B Hegde[1] and B.Sathish Babu[2]

[1,2] Department of Computer Science and Engineering,
Siddaganga Institute of Technology, Tumkur, Karnataka, India.
[3] Pallapa Venkatarm,
PET Unit, Electrical Communication Engineering,
Indian Institute of Science, Bangalore, Karnataka, India.

## ABSTRACT

*In Manet's Dynamics and Robustness are the key features of the nodes and are governed by several routing protocols such as AODV, DSR and so on. However in the network the growing resource demand leads to resource scarcity. The Node Mobility often leads to the link breakages and high routing overhead decreasing the stability and reliability of the network connectivity. In this context, the paper proposes a novel opportunistic AODV routing scheme which implements a cognitive agent based intelligent technique to set up a stable connectivity over the Manet. The Scheme computes the routing metric (rf) based on the collaboration sensitivity levels of the nodes obtained based through the knowledge-based decision. This Routing Metric is subsequently used to set up the stable path for network connectivity. Thus minimizes the route overhead and increases the stability of the path. The Performance evaluation is conducted in comparison with the AODV and sleep AODV routing protocol and validated.*

## 1. INTRODUCTION

The information diffusion dynamics has become seamless with the increased use of pervasive computing devices. The MANET is one among the important technologies supporting pervasive computing by enhancing the mobility in the dynamic environment. MANET's [1] are distributed, self-configurable, infrastructure less Ad hoc class of wireless networks with no fixed apparatus. MANET's find applications in many areas ranging from commercial to critical as well from consumer to the military. MANET's diversified use is due to its features such as robustness to cope up with dynamic network topology; fluctuating, high bit error rate and various links; collaboration among the nodes and so on [2]. The typical applications are in dynamic unplanned situations [3] requiring emergency management such as recovery from disasters; an earthquake, flood, chemical accidents and so on. MANET's can handle the volatile topology [4] through dynamic routes. It has hosted routing protocols [5] such as AODV, DSR, DSDV, and TORA and so on. Even though an investigation says no specific protocol is sufficient for all the scenarios, AODV [6] is the popular imploded on-demand reactive, table driven protocol.

AODV consists of two phases of routing, i.e., route discovery and route maintenance. AODV discovers routes when required and are maintained just as long as needed. When a node wishes to send a packet to some destination, on the unavailability of the current path initiates the route discovery phase by broadcasting route request (RREQ) packet. On comparison, AODV outperforms DSR protocol [7] [8] in stressful situations having increased mobility and traffic due to fewer delay requirements.

However, the AODV protocol exhibits several pitfalls that need to overcome for better connectivity during the information diffusion. The major limitation of the protocol is the routing overhead [9]. It only maintains the link information between the communication nodes. Link failures will be more at higher mobility, triggering more route request packets, and thus the network has to suffer from the high frequency of route discovery phases which degrades the performance. Mostly, these link breakages are due to lack of resources such as energy, memory, and width, computing power and the mobility of nodes.

There is a need for a novel scheme that minimizes the route request overhead by identifying the resource requirement and takes knowledge based decision prior the link breakage. An efficient resource pooling technique is required to identify, manage and track the resource on time. Thus the probability of link breakage can be determined and proactively notified to the mobile nodes. Using the information the node needs to take an intelligent decision towards maintaining the connectivity without triggering the route re-discovery phase. But due to the dynamic mobility of the nodes, getting the resourceful node for computation and connectivity will be difficult with conventional AODV scheme, as it requires an end-to-end connectivity between the source and destination. Hence a computing paradigm is needed that makes use of the intermittent connectivity within the network based on opportunistic contacts between nodes. The paper proposes a novel intelligent scheme to reduce the routing overheads of the AODV scheme using opportunistic computing technique based on cognitive theory.

## 1.1 OPPORTUNISTIC COMPUTATION

An opportunistic computing is a kind of pervasive computing [10], which takes the benefits of randomness and uncertainty in the environment to establish the communication opportunistically between pairs of diverse devices and applications. The opportunistic computation takes place in the nodes that are part of an opportunistic network.

## 1.2 OPPORTUNISTIC NETWORK (ON)

It is the class of MANET's which exploits the contact opportunity among the nodes for the communication between the source and destination without the end-to-end connectivity [11]. The favorable circumstances such as interconnectivity, context-activity and so on, create the contact opportunities.

*Formulation of opportunistic network:*

Let us consider a graph of MANET, G{V, E}, with 'V' number of nodes and 'E' wireless links among them as shown in figure 1.

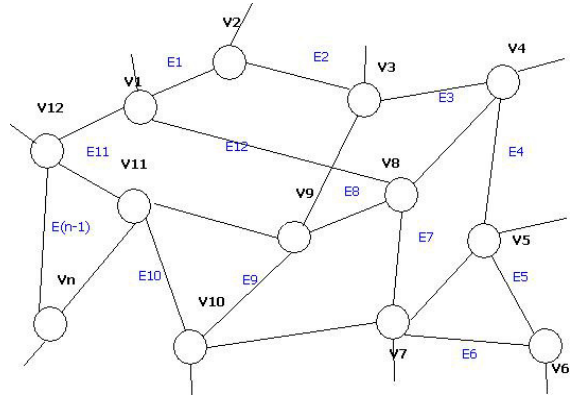

Figure 1: MANET set represented by graph G {V, E} with 'n' number of V and 'n-1' number of E.

Then H{P, L} be the subgraph of G with P ={ $n_k$} number of nodes and L={ $l_{i,j}$}, where i≠j wireless links as shown in figure 2, where S represents the opportunistic network area.

Then H is called an opportunistic network over a Euclidean plane 'S' iff the affinity of opportunistic contact between the nodes $n_i \in P$ and $n_j \in P$ is greater than the average degree of opportunistic contact ($\Delta^{avg}$) between the two nodes as shown in figure 2.

i.e., $H\{P,L\} = \begin{cases} \emptyset & if \quad P=L=\{\} \\ (n_i, n_j) \cup S & if \quad \Delta(n_i, n_j) \geq \Delta^{avg} \end{cases}$

$\Delta$ denotes the degree of opportunistic contact between the nodes and $\Delta(n_i, n_j) = \Delta|n_i| - \Delta|n_j|$. The degree of opportunistic contact of node $n_i$ w.r.t its neighboring nodes (k) is, $\Delta n_i = \sum_{k=1}^{m} \Delta n_{i,k}$. Similarly for $n_j$, $\Delta n_j = \sum_{k=1}^{m} n_{j,k}$.

Let D {N, W} the weighted contact graph with N number of neighbor nodes and W weight matrix= {$w_{ij}$} be the sub graph of H. Weight matrix indicates the relationship strength between the nodes i and j. The weight $w_{ij}$ is a correlated matrix of the parameters contact frequency $f_{ij}$ and the sum of all contact durations $t_{ij}$.

$$w_{ij} = \left( \frac{f_{ij}}{\sigma_f}, \frac{t_{ij}}{\sigma_t} \right)$$

Where, $\sigma_f$ and $\sigma_t$ are empirical standard deviations and i, j ∈ N.

The $w_{ij}$ is transformed to a scalar value $W_{ij}$ using the principal component.

The average degree of opportunistic contact $\Delta^{avg} = \frac{W_{ij}}{N}$, N is the number of connected neighbors.

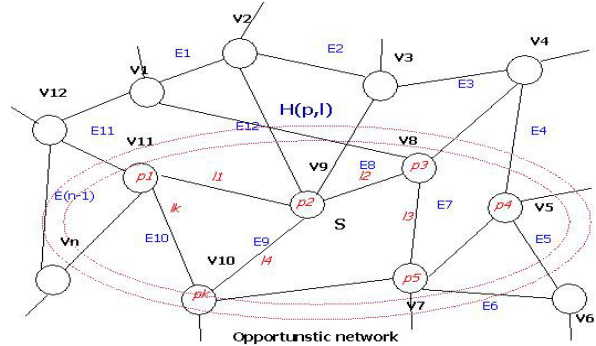

Figure 2: Formation of an Opportunistic network represented by sub graph H (P, L) within the MANET graph G (V, E)

### 1.3 OPPORTUNISTIC CONTACT (OC)

The nodes which are within the reachable region are said to be in opportunistic contact if they share any one or all of the characteristics such as the common goal, shared context activity, same background, common destination and so on. The degree of opportunistic contact ($\Delta$) indicates the affinity between the nodes to have an opportunistic contact.

Therefore the Opportunistic contact between the nodes $n_i$ and $n_j$ is $OC\{n_i, n_j\} = \max \bigcup_{i=0}^{k} \{\Delta_i\}$

where, the degree of opportunistic contact $\Delta = \dfrac{Number\ of\ nodes\ having\ the\ affinity\ \text{for OC}}{Total\ number\ of\ nodes}$

The number of nodes having affinity for OC is given by, $nrc_\pi \sum_{k=1}^{n} a(n_k)$,

where a(n) represents the affinity of node n towards the OC and is given by,

$$a(n) = E \mapsto \bigcup_{S \in E} \{Cn \cup Rn \cup Ln\}$$

E denotes the MANET environment with $s \subseteq E$.
Let $C_n$ be the set of context activities in E at time NT.
Let $R_n$ be the set of resource available in E at time NT.
Let $L_n$ be the set of nodes towards same destination in E at time NT.

The $\Delta$ is parameterized using the metric called *Opportunistic Contact level (OCL)*. The various context activities at the $K^{th}$ node will choose the OCL into any one of the 3 set levels within range of {0-2}, where 0 is the highest degree and 2 is the least.

The $j^{th}$ context activity of the $K^{th}$ node is given by, $C_j(k) = (l_{ij}^{(k)}, x_{ij}^{(K)})$

where $l_{ij}$ is the link between the nodes $n_i$ and $n_j$. $x_{ij}$ is the context similarity metric between the nodes $n_i$ and $n_j$. Similarly, $R_j(k)$ and $L_j(k)$ can be computed.

The $x_{ij}$ is a function of the three context parameters. The age of last context match ($A_m$), which is obtained from the timestamp attached with each context match. The number of context matches between the nodes ($N_m$) obtained from log history. The location traces in the history log gives the similar context location (Lm).

Let $\{Tr_i\}$ be the set of location traces for node $n_i$ and $\{Tr_j\}$ be the set of location traces for node $n_j$ then,

$$L_m = \frac{|Tr_i \cap Tr_j|}{|Tr_i \cup Tr_j|}$$

The contact similarity metric $x_{ij}^{(k)}$ is given by,

$$\begin{cases} x_{ij}^{(k)} \geq 0 & \forall\ j=1,2,....n \\ if\ 0 \leq x_{ij}^{(k)} \leq 0.5\ \forall\ k \in\ E,\ then\ \varepsilon \\ if\ 0.5 \leq x_{ij}^{(k)} \leq 1\ \forall\ k \in\ E\ then\ \delta \end{cases}$$

Where $\varepsilon$ denotes the Vagrant context similarity and $\delta$ denotes the maximum context similarity.

The computation performed on nodes by opportunistically availing the other resources of the environment, along with resources available on its device within the opportunistic network is called *opportunistic computation*. Here each node exploits the transient unpredicted contacts to cooperate and share each other's content, resources, and services [11, 12] (M. Conti, 2010).

### 1.4 COGNITIVE THEORY

Cognitive Agent [12, 13], is an intelligent software agent. It is an autonomous and responsive entity which reflects the state of information of the environment which is represented using cognitive terms. Similar to human perceptual experience and understanding of the surrounding environment, CA's can provide an immediate or a controlled response to the changes in the environment [14, 15]. Cognitive agents have self-learning, self-improving and self-instructible ability [17, 18] and therefore, reduce the network overhead. Cognitive agents are suitable for adaptive, dynamic, context-based behaviourism, decision making, human mobile network interaction system [16, 19, and 20] and so on.

### 1.5 PROPOSED COAODV

The proposed Cognitive agents based Opportunistic AODV (COAODV) scheme uses Mobile cognitive agent (MCA). The mobile nodes in the network embed the MCA with them. The MCA is having communication with all reachable MCA's track the availability of resources through the Behavior-Observation-Belief (BOB) cognitive model (b s bsbu et al., 2009) and generates the recommendations about resource availability based on the routing factor. Upon the recommendation, the AODV routing protocol will opportunistically route the messages towards the destination before losing the current connectivity and avoids the triggering of the route discovery process.

*Organization of the paper:* The rest of the paper is organized as follows, Section 2 gives some of the related works, Section 3 discusses the proposed scheme, section 4 illustrates the results, and finally, Section 5 draws conclusions.

## 2. RELATED WORKS

Some of the research works conducted in AODV routing technique to discuss the pitfalls of the protocol are mentioned here.

Sung-Ju Lee et al., (2000) proposed a scheme to keep the alternate backup paths during link establishment to avoid failures without using additional control messages. The mesh structure uses an alternative route table with the alternate route utilization mechanism like DSR protocol and tested the performance of AODV protocol. The mechanism increased the robustness in mobility but degraded at high traffic load due to multiple collisions.

Naif Alsharabi et al., (2005) proposed a Packet Received Time (PRT) based prediction technique in AODV route maintenance phase by measuring the power of the received packets to predict the link breakage in the priory and rebuild the route. The simulation results show that AODV_PRT delivers more packets at higher mobility, but the end-to-end delay will be more than the original AODV and remains same as original at lower mobility as well the control overhead will be more.
C. T. Cuong et al., (2014), Proposed routing algorithm MAR-AODV (Mobile Agent- AODV) to avoid the link breakage due to congestion using mobile agents which are added to AODV for updating traffic density at each node, thereby improving the network performance.

Xin Ming Zhang et al., (2013), proposes a neighbor coverage-based probabilistic rebroadcast (NCPR) protocol, to reduce the routing overhead through avoiding the frequent link breakages. It exploits the dynamically calculated rebroadcast delay metric, rebroadcast probability and the neighbor coverage knowledge got from 1-hop neighborhood information. The performance evaluation shows it reduces the redundant packet retransmissions and alleviates routing overhead. But, the protocol has considered only the packet collision not connectivity loss due to the random change in the topology.

D.G. Reinaa et al., (2012), have shown the importance of MANET's in the disaster management scenario. It proposes a computational approach using reach ability metric to optimize the connectivity through modeling the MANET's mobility during disaster using a genetic algorithm for both Intra and intercommunication.
Tahar abbes Mounir et al., (2013), propose sleep-AODV protocol in the disaster scenario to keep the connectivity during the communication. Based on the mobility of the nodes the energy consumption rate is divided into three periods and limits excessive routing messages. However, the performance evaluation shows the degradation in the functionality under high-level traffic due to more alternate paths.

C. A Ntuen et al., (2011), give a framework to simulate the MANET's network as an intelligent agent using the cognitive and behavior based modeling in the battlefield scenario. The proposed scheme uses an OODA (Observe, Orient, Decide, Act) model to represent the MANET's intelligent behavior. The paper provides a foundation for modeling agent behaviors in a way plausible on human behavior.

## 3. PROPOSED COGNITIVE AGENTS BASED OPPORTUNISTIC AODV ROUTING SCHEME (COAODV)

The COAODV, like AODV is a reactive on demand scheme. It has both route discovery and route maintenance phase. The difference is, the path is set up only with those nodes having the resource availability level more than the required threshold so that the link breakage due to resource starvation is not found and reduce the often triggering of the route discovery phase. Once the path is set up, the message will be forwarded opportunistically towards the destination (Without the End-to-End connection between source and destination) and reduce the dropping of packets.

### 3.1. COAODV scheme architecture

This section gives the architecture of the COAODV scheme by discussing its functional components and their working.

The design of COAODV consists of a mobile cognitive agent on the node along with the Contact Analyzer (CA) and the Resource Record (RR) as shown in figure 3.

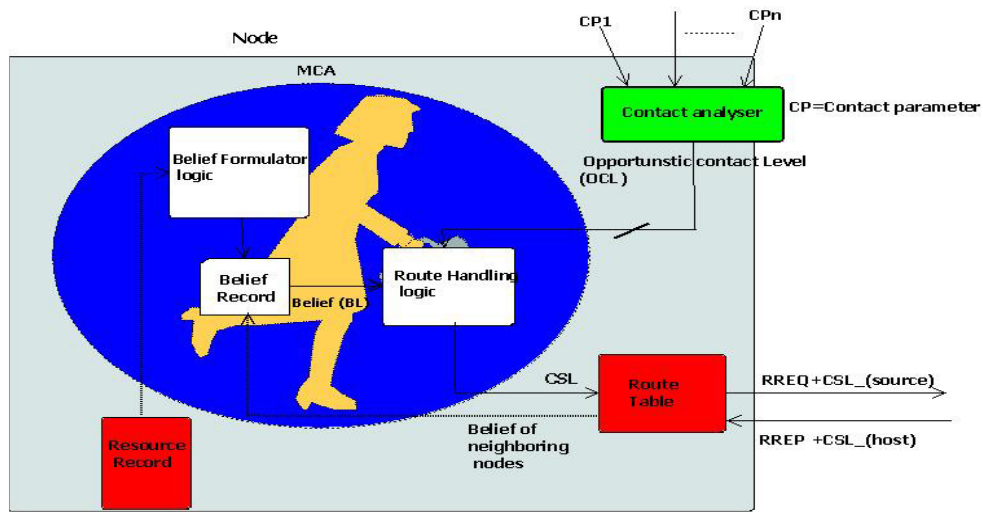

Figure 3. Architecture of the COAODV

The contact analyzer classifies the external parameters obtained from the surrounding environment and group the close contact nodes to different levels of opportunistic contact based on their level of sensitivity towards the formation of opportunistic contact between the nodes. The Resource Record (RR) is a tuple stores the resource information such as <Type, ID, total> connected to the node where Type signifies the resource type, ID gives its unique identification and total give the number of such resources.

The main components of MCA include the belief formulator logic, belief record, route handling logic and the opportunistic contact level obtained from contact analyzer. MCA's of a node will be monitoring the resource availability of the host node as well as stores the resource capabilities of the neighbouring nodes in the form of beliefs. The belief formulator logic generates these beliefs and stores in the belief record. The route handling logic will initiate a route discovery by flooding Route Request (RREQ) packets.

The opportunistic contact level and the belief generated together will set the parameter called collaboration sensitivity level (CSL); *The ability of the node to collaborate with the host node to coordinate the communication.* Using the CSL the reliability of the connection is computed using the parameter called routing factor (RF) to set up the path.

### 3.2 Components of COAODV

The COAODV architecture consists of the following components.

### 3.2.1 Contact Analyzer

CA will analyze the sensitivity of the incoming transactions called contact parameters (CP) to find whether the close contact node can form an opportunistic contact with the node. It extracts the multiset of context activities 'C $_i$ 'having the Multiplicity m (c) from the CP and compute the OCL as shown in algorithm 1. The obtained OCL will be concatenated with the beliefs got from the belief record to get a CSL value.

Algorithm 1: Working of contact analyzer

```
1: Begin
2: Initialize the session
3: Initialize the Log.
4: INPUT: Contact parameters CP;
5: OUTPUT: Opportunistic contact levels grouped into different groups.
6: Set multiset S to ø, contains the set of context-activity  S = [C_1^m, C_2^n, ........ C_n^t] each occurring  m, n· · · · ·t
number of times.
 7: while Not end of session do
8: Capture the contact parameters CP externally and from log.
9: for each time multiset S≤ n  do
10: Extract the context activity C_n to multiset  S.
11: End for
12: Set multiset Q to ø, contains the set of resource availability
13: Q = [R_1^m, R_2^n, ........ R_n^t] each occurring  m, n· · · · ·t number of times.
14: End for
15: Set multiset V to ø, contains the set of node links  V = [L_1^m, L_2^n, ........ L_n^t] each occurring  m, n· · · · ·t
number of times.
 16: End for
17: Set T to ø, set of all the multisets, i.e.,  T=[s, Q, V]
19:  compute the various combination of contact opportunities  $\sum_{n=0}^{i} O(T,n) = \frac{T}{n!(T-n)!} = 2^i$
13: Based on the intensity of contact  $2^i$ output will group in to different opportunistic contact levels (OCL) ranging from 0 to 2
14: end while
15: Return OCL
16: Periodically refresh Log
17: End
```

### 3.2.2 Belief Formulator Logic

The logic is based on the BOB (Behavior-Observation-Belief) model [B S babu, 2009]. The three constructs namely Behaviors identifier, Observations generator, and Beliefs formulator are put together to form a Belief formulator logic as shown in figure 4.

The belief formulator will formulate the belief over the availability of resources on a node in an ongoing session. The logic makes use of three essential knowledge quantifiers called Behaviour (The actions or reactions of nodes while formulating and executing the transactions), Observation (The summarization of various behaviors exhibited by the mobile nodes during transaction execution) and Belief (Representation of information about the world or an entity from the obtained observation perception). The BOB model will capture the behaviour parameters from external environment from which it deduce the observations and generate the beliefs. It works as in algorithm 2.

The belief record stores the belief. The beliefs generated are classified based on the availability of resource in the host node. The beliefs are grouped into any one of the four classes, *Casual, Slack, Patron or Vargant based* on the rate of mobility and resource availability of the host node as shown in Table 1. The beliefs need not be mutually exclusive.

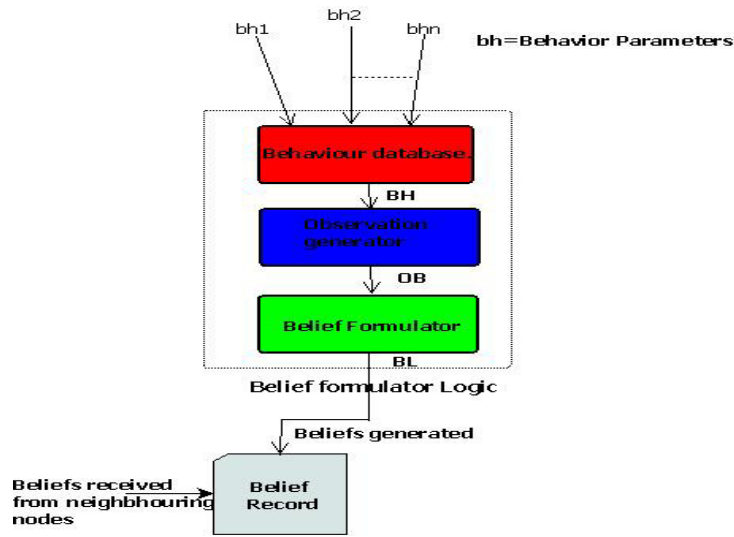

Figure 4: Belief formulation logic

Algorithm 2: working of Belief formulator

```
1: Begin
2: Initialize the session
3: Initialize the Log.
4: INPUT: behavior parameters bh;
5: Intermediate outputs: Behavior (BH), Observation (OB)
5: OUTPUT: Generated belief BL.
6: Set the belief set BF to ø, contains the set of beliefs generated in belief formulator
7: initialize Behavior Identifier (BI)
8: while Not end of BI do
9: Capture the behavior parameters bh externally and from log.
10: Compute the weightage α_{bhk} for bh_k
11: α_{bhk} = f(bh_k; bh_l;m)
12: BH_i = f(α_{bhk} ;Current value of (bh_k);Max: possible value of(bh_k))
13: end while
14: Behavior set BP=BP U BH_i
15: initialize Observation generator (OG) with
16:for all observation obi , select favorable BH_j
17: while set OB ≥ 0 do
18: OB = obi U OB_i occurred most no. of times during the session.
19: end while
20: Initialize the Belief generator (BG)
21: Pass OB to BG.
22: for all BLi€ BL, select favorable OB_j
23: BL_i = BL_i U OB_j
24: while BL ≥ 0 do
25: BL = BL U BL_i
26: end while
27: Return BL
28: Periodically refresh Log
29: End
```

Table 1: Classes of Beliefs

| Class | Characteristic |
|---|---|
| Patron | *High profile* i.e., capable to give large amount of resources with *High mobility* node |
| Casual | *Low profile* i.e., capable to offer small amount of resources with *High mobility* node |
| Slack | high profile with low mobility node |
| Vargant | low profile with low mobility node |

Belief Record is a tuple <Node ID; Belief generated> which stores the generated beliefs with the corresponding node ID. The belief generated by the MCA running on the host node will be stored as self-belief. Beliefs obtained from all the reachable neighbouring nodes are stored along with their node ID.

These beliefs with the same collaboration sensitivity level are exchanged among reachable nodes on demand to make the decision about having collaboration with those nodes when required in the form of resource request and reply.

### 3.2.3 Route handling Logic

The logic obtains the Belief (BL) and the Opportunistic contact level (OCL) as the inputs.

International Journal of Ad hoc, Sensor & Ubiquitous Computing (IJASUC) Vol.8, No.1/2/3, June 2017

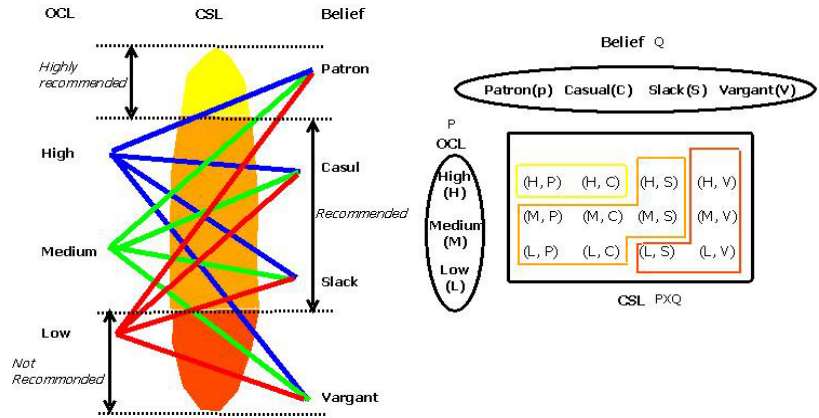

Figure 5: The CSL derived from Cartesian product of OCL and Belief

It is responsible to set the nodes collaboration sensitivity level (CSL) of the host node. The Cartesian product of the two inputs will generate the CSL as shown in figure 5. The CSL can be of levels *Highly recommended (2), Recommended (1) or Not recommended (0)* based on the Cartesian parameters. The CSL is derived from the two inputs OCL and BL as in algorithm 3.

Algorithm 3: working of Route handling logic

```
1: Begin
2: INPUT: Belief BL, Opportunistic contact OCL;
3: OUTPUT: Collaboration sensitivity level CSL;
4: initialize the parameters P and Q with 0
5: while P ≤ 2 do
6: call the contact analyzer and extract the value of the OCL to OCL P
7: end while
8: while Q ≤ 3 do
9: call the belief formulator and extract the value of BL to BL_Q
10: end while
11: for all (OCL _P , BL_Q) do
12: obtain I = ∏(OCL_P × BL_Q), Where '×' represents the Cartesian product of set OCL and BL
13: end for
14: If I == <(H,P)||(H,C)> then CSL ⇐ Highly recommended
15: If I== <(H,V)||(M,V)||(L,S)||(L,V) > then CSL ⇐ Not recommended else CSL⇐Recommended
16: Return CSL
16:  End
```

### 3.2.4 Route table

The route table is a tuple; it stores the information required to assist the routing of the packets towards the node having the better stability such as CSL value. The node initiating the routing to transfer the message is called as *Source node*; The intermediate nodes that assist the source node to set up the path are called *Host nodes*; The intended node to transfer the message is called as *Target node*. The fields of the routing table are as shown in Table 2.

Table 2: Route table of COAODV

| UID | CSL | Source UID | Source CSL | Target UID | RR | $RF_s$ |
|-----|-----|------------|------------|------------|------|----|
| 01  | 2   | 00         | 2          | 03         | $I_1$-D | 2  |
| 00  | 2   | 00         | 2          | 03         | S-$I_1$ | 1  |

Where, UID = The unique number of a host node.
CSL= Collaboration sensitivity level of the host node.
Source UID = The unique number of a source node.
Source CSL = The collaboration sensitivity level of source node
Route record (RR) = The forward path the packet may traverse.
Destination node = Target node for data packet.
Routing factor ($RF_s$) = Indicates the connectivity reliability of the route on source.

### 3.3. Routing Procedure

When a node initiates the data routing, on the unavailability of the route, the route discovery phase starts. During the route discovery phase, the source node clones the MCA along with the RREQ packet to all its neighbor nodes as in figure 6.

| IP Field | options field | option length | Source ID | Target address | CSL |

Figure 6: Route request packet

The belief formulator of the cloned MCA will generate the belief about the resource availability on the node and sends it to route handling logic.

The route handling logic extracts the OCL from the contact analyzer of the node over which it resides. It will generate the CSL of the host node and copies to the RREP packet. The RREP packet will be sent back to the source node.

Several RREP packets will reach the source node and updates its routing table. The MCA of the source node will compute the Routing factor (RF) based on the established CSL value and stores in the route table.

$$RF = min\ (CSL_{source}, CSL_{host})/CSL_{source}$$

Where, $CSL_{source}$ = The collaboration sensitivity level of the computing node.
$CSL_{host}$ = The collaboration sensitivity level of the neighboring host considered.

The source node will select only the path which has routing factor greater than 1 and initiates the data transfer as shown in figure 7.

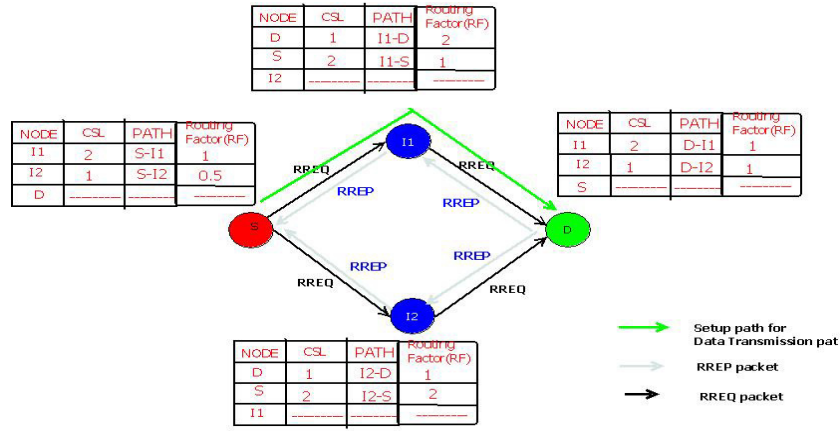

Figure 7: Route path setup in COAODV

The belief formulation is a proactive process running on the MCA and each time the node allocate the resources the belief over the source node changes and thereby the CSL of the node varies. MCA's on each node will be monitoring its level and during the data packet transmission, with decreasing CSL (long before the completion of N/2 packets) the MCA Proactively stimulate on demand belief formulation process and updates the belief record. Again the route handling logic will verify its CSL with the requested CSL level and on no match computes the routing factor. If required (if it's RF<1) handovers the routing responsibility to a neighbouring node having RF>1.

As in figure 8, initially with sparse nodes being isolated the degree of opportunistic contact will be less. With the increase in node density in both with and without agents scenario the Δ increases. However with further increase in the node density, the due to resource starvation Δ decreases in network without agents. As the resource availability is predicted and the route is setup based on recommendations Δ increases with increase in node density.

The plot in figure 9 gives the rate of generation of different classes of collaboration sensitivity levels. In can be observed that the class of *'Recommended'* will be more compared to *'Highly recommended'* and *'Not recommended'* signifying that a controlled recommendations are given so a reliable path is set up for routing and reduces the frequent triggering of route discovery phase.

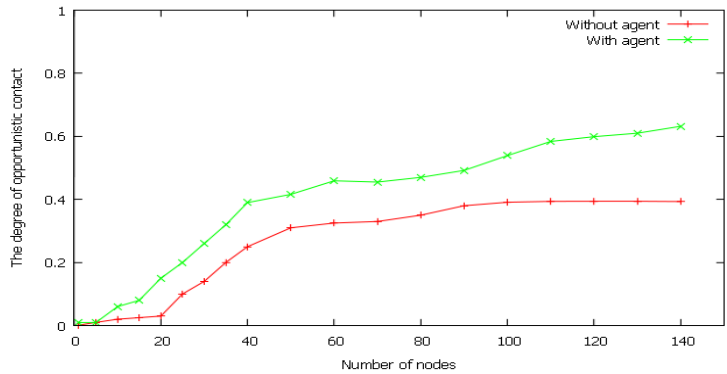

Figure 8: Number of nodes Vs degree of Opportunistic contact (Δ)

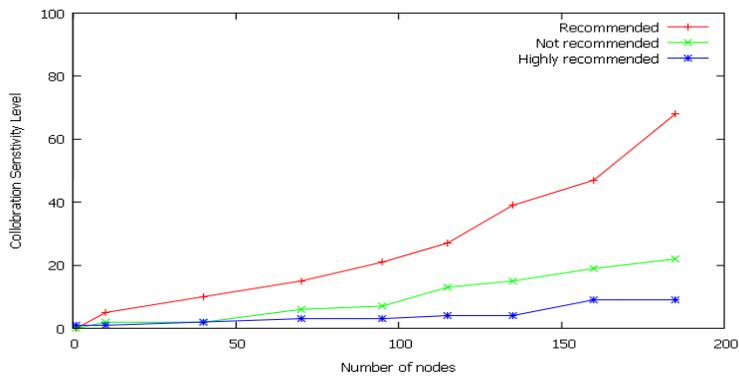

Figure 9: Number of nodes Vs CSL

## 4. VALIDATION OF COAODV

The COAODV is a stable routing protocol for MANET's where the stable route is obtained between the source and destination based on the routing factor of the intermediate nodes which depends on the resource availability for the data transfer. The protocol is validated through comparison of the proposed protocol with a sleep-AODV protocol [26].

### 4.1 Sleep-AODV

It is an extended power aware routing protocol, which takes into account the behaviour of the node (stable, unstable or isolated) to maintain the connectivity. The purpose of the sleep-AODV protocol is to limit excessive routing messages caused by a large number of neighbours because a routing table with many entrances involves a significant number of control packets, which consumes more energy spent in the transmission, reception, and overhearing mechanism.

#### 4.1.1 The principle of Sleep-AODV protocol

In Sleep-AODV protocol a node switch to standby mode for some time according to the network status as shown in figure 10. The general operation of a node is represented by four states idle (initial stat), Sleep, receive or transmit. Each node goes to sleep for a fixed duty cycle called sleep time by turning off the radio. Then wakes up and listens in the idle state to see if any node initiates communication either through send or receive state.

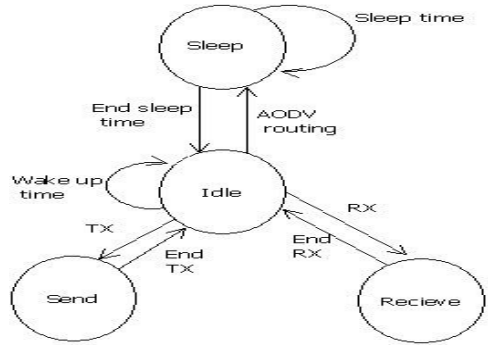

Fig 10: State diagram of Sleep-AODV protocol

Sleep-AODV protocol is designed to detect the unstable node, isolated nodes and in active paths. On detecting such conditions it turns off the radio and reduces the energy resource consumption. Hence, Sleep-AODV protocol avoids the reroute discovery due to energy deficient link failure by reducing the passive listening period of the nodes.

### 4.2 Experimental Setup

An experimental setup is built to analyze the behaviour of the COAODV protocol in comparison with Sleep-AODV. The scenario is setup for post-disaster management equipped with mobile devices, in and around the disaster area. Immediately after the occurrence of the disaster, the disaster relief operation will takeoff, and the rescue team will start the search and rescue activity. The members of the rescue team will be working by grouping themselves into many subgroups. Possible communication subnets:

- To have a personal connection between them, the team members will carry the mobile devices such as smart phones, tablets, palm top, laptops and other devices having the wireless capabilities. They can form the subnets in and around the incident location( subnets created in and around the incident location due to the usage of computational devices bundled with several wireless network interfaces).

- To clear the casualties and to transport the rescued victims for the medical supervision, rescue vehicles will be moving throughout the incident area and can form the vehicular network (The vehicular network created by the rescue vehicles moving around with the devices come in contact with it on both the sides of its path).

Thus, the application scenario depicts that in spite of collapsed communication network there are an intermittent group of subnets available with a various level of wireless capabilities and device characteristics, which can be made use to setup a communication network among the personnel of the rescue team.

Let us consider the scenario in which pooling of resources is done in a completely or partially collapsed communication infrastructure formed amid the affected area, as shown in Figure 11. It involves a mix of mobile devices with random mobility patterns and sporadic data transmission rates having various computational capabilities. For example, the team of rescue members (nodes) holding the wireless communication devices, such as smart phones, laptops, and so on, will move with a group and have slower unpredictable movement. Thus, there will be much delay in message passing. The rescue vehicles (nodes) will have faster and slightly predictable path, leading to faster message transmissions and so on.

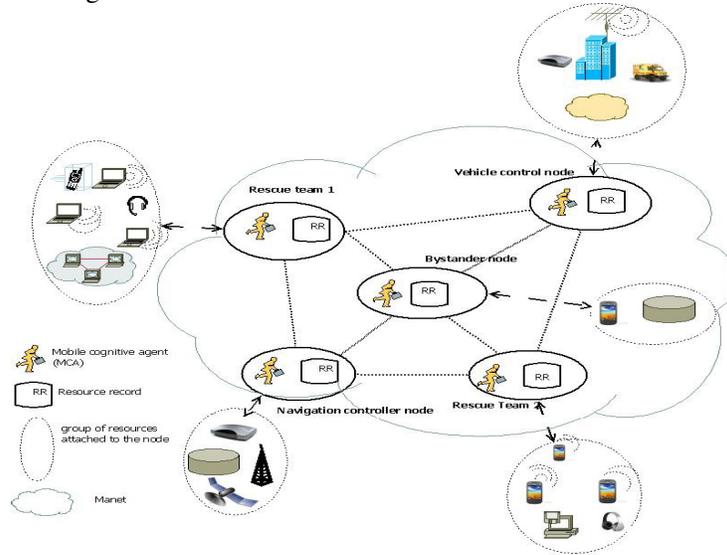

Figure 11: A Post disaster resource pooled MANET's communication scenario

Initially, these mobile resources are registered into any one of the following four groups based on their resource capacity and skills as shown in table 3.

Table 3: Types of mobile nodes

| Node | Description |
| --- | --- |
| Rescue Team | A group of trained people allocated with pre-registered devices for faster searching and triage. |
| Bystander | The local volunteer individuals registered to assist the rescue team |
| Navigation Controller | Registers the new devices and guides the search and triage operation |
| Vehicle Controller | Tracks and coordinates the rescue vehicles used for victims triage |

The model implemented with the COAODV has events such as *registration*, *quantification*, *mapping*, and *belief generation*.

The MCA of the navigation controller, being the registration authority, will clone itself on the newly registered device with the necessary initial handshaking. The copy of the agent from the registration authority will migrate onto the new device, ready to capture the behaviours and formulate the beliefs about resource availability. The MCA on the new device (node) will be available for communication with the MCAs of the other nodes.

## 4.3 Test bed description

The evaluation of COAODV is done on a Riverbed opnet [28] test bed for a post-disaster scenario as shown in figure 12. The simulation environment is considered in an area of 500 x 500 unit, where mobile nodes are distributed randomly. The environment comprises 200 wireless mobile nodes embedded with the MCAs designed using agent factory, based on 802.11b configured in ad-hoc AODV routing mode.

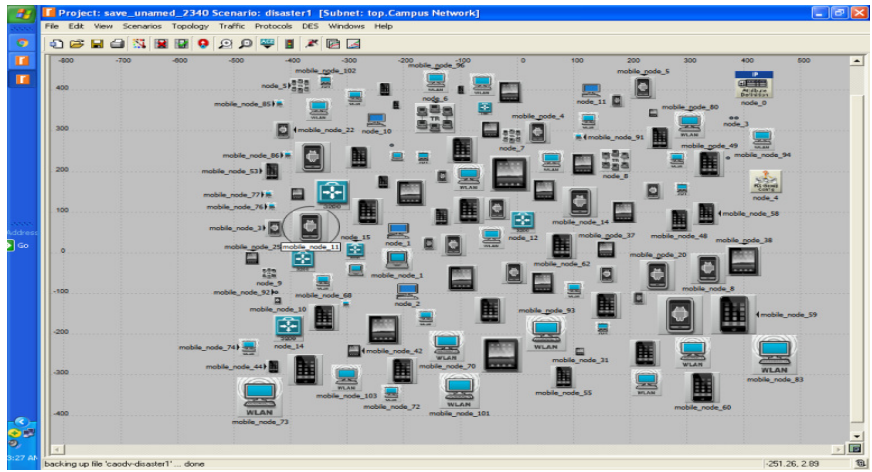

Fig 12. The simulated Test bed

The nodes' mobility is configured to a random waypoint with a communication range of 60 m and pause time ranging from 0 to 150 ms. The nodes are running with the Linux kernel on the IP networking protocol stack that is essential in studying the scenario, of which about two-thirds are mobile nodes such as a WLAN iPad, Android mobile, WLan iPhone, MANET workstations, WLAN workstations and so on supporting 1Mbps WLAN with CBR traffic. The nodes belonging to the rescue team and bystanders have a reduced communication range of 10–30 m. The remaining nodes are static for all transmission range such as survived infrastructure, vehicle controllers, and navigation controllers emulated through Mac workstations with wireless cards. The registered devices will attach to the node's resource pool and be mapped with its class of resource and will then update the RR. The test bed gives various possibilities of a node coming in opportunistic contact (nodes in the same direction, toward same locations, executing the same task) with other registered nodes for resource pooling. It uses the RFC 2501 [27] mobile IP networking wireless protocol for node interaction.

The performance metrics used for the comparison and analyses of the protocol are as follows:

*Number of control packets*    The amount of control packets *generated*/connection for route discovery and rediscovery.

*Packet delivery rate (PDR)*    The ratio of no. of packets sent to the no. of packets received at destination.

*Packet loss*    The number of CBR connection between the nodes to the packet loss.

*Overhead traffic rate(OTR)*     The ratio of the control traffic to the total traffic
                                 generated over the number of connections.

*End-to-End delay*               The average time between the start of packet transmission to the
                                 time it reaches the     destination with traffic density. The
                                 breakdown time includes the delay due to route discovery
                                 process and the queue in data packet transmission.

Initially, to establish the path between the nodes the route discovery phase is initiated. The RREP packets are flooded over the neighbour nodes whose density increases with increase in connection over the network as shown in figure 14. In sleep-AODV, with an increase in the connection the control packet flooding is reduced by 20% as the nodes will go to sleep in a specific duty cycle. In COAODV, control packets cloned with the MCA are flooded only during the initial route discovery phase. The MCA's will compute the belief on connectivity and reduce their discovery phase. Due to this with an increase in the number of connections flooding of control packets is been greatly reduced.

Figure 14, shows the rate of overhead traffic with the varying number of connections. At the lower OTH, all the protocol behaves similarly. With an increase in the number of connections, AODV suffers from highest OTR due to more rediscovery phases because of a number of collisions leading to link breakages and more RREP packet broadcast. Sleep-AODV will detect and separates unstable, isolated nodes and inactive path. So it will exhibit a reduced OTR compared to AODV protocol. COAODV exhibits the minimum OTR because of least number of link failures. COAODV will predetermine and the will setup the path only with the links having required resources. It maintains the link with the obtained opportunistic contact and thereby provides higher link stability.

Figure 15 gives the plot of Packet delivery rate which is highest in COAODV protocol. In COAODV protocol, the packets will be sent only through the path which is having greater stability and thereby reducing the number of packet retransmission. In AODV repeated link breakages with the increase in a number of nodes have lesser PDR. Initially, Sleep-Aodv has least PDR due to the number of inaccessible nodes and increases with increase in node density.

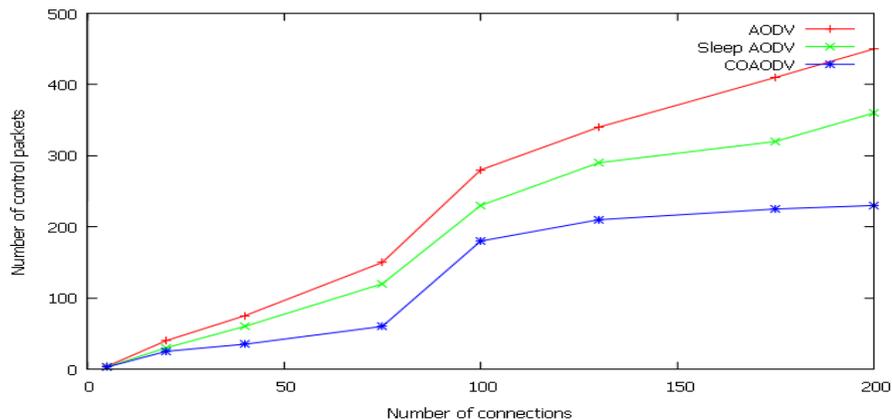

Fig 13. Number of CBR connections Vs Number of control packets generated

International Journal of Ad hoc, Sensor & Ubiquitous Computing (IJASUC) Vol.8, No.1/2/3, June 2017

Figure 16 shows the plot of a number of connections Vs End to end delay. The Sleep AODV protocol has a large delay, as most the time the nodes will be in sleep mode and are inaccessible. The COAODV has least End to end delay due to the fact that the path will be established only among the nodes having the routing factor greater than the threshold level and hence provides greater stability and reliability.

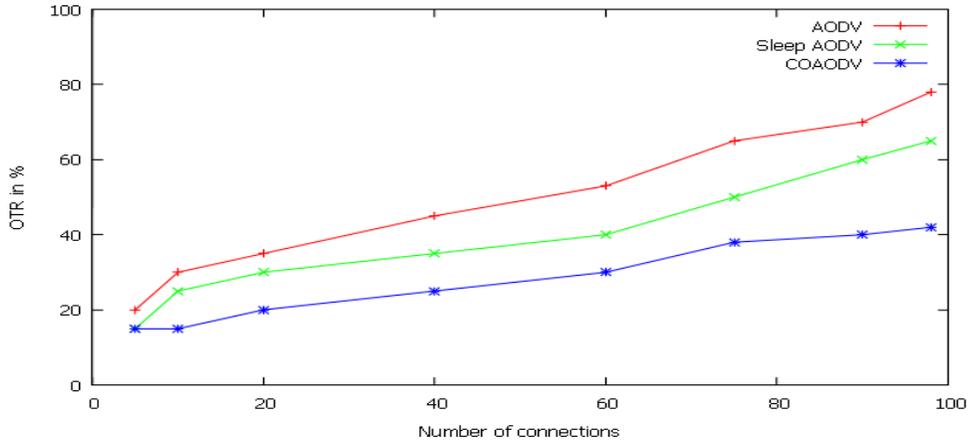

Fig. 14 Number of connections Vs OTR

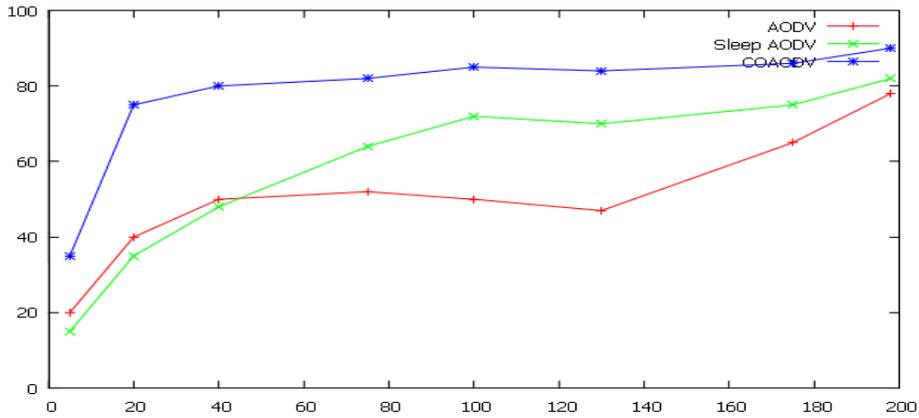

Fig 15. Number of nodes VS PDR

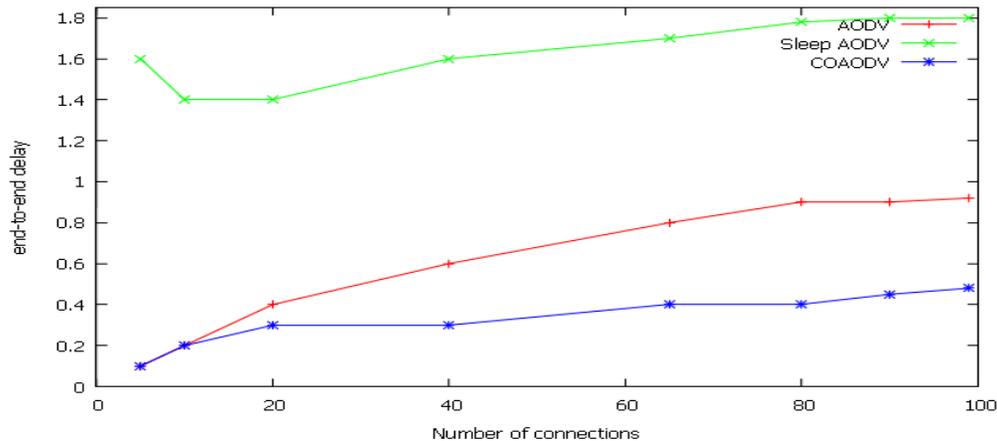

Fig 16. Number of connections Vs E-E delay

## CONCLUSION

The Opportunistic computation in MANET's exploits the contact opportunities to establish the communication among the nodes without end- to-end connectivity. The OC with cognitive agents can provide a better decision on route set up reducing the unnecessary route discovery phases. The paper presents a novel COAODV scheme to improve the stability and reliability of the network connectivity in MANET's. The scheme uses CA-based OC principles to determine the resource based routing stability of the path. The technique is implemented based on the BOB model for a disaster management scenario. The scheme used works close to human cognitive behaviour in determining the stable route reducing the frequent route discovery phase and is therefore an efficient approach for MANET's. The scheme on comparison with the AODV and Sleep-AODV provides a reliable and adaptive and stable routing in a disconnected environment.